# Inverse Spin Hall Effect in nanometer-thick YIG/Pt system

O. d'Allivy Kelly[1], A. Anane[1*], R. Bernard[1], J. Ben Youssef[2], C. Hahn[3], A-H. Molpeceres[1], C. Carrétéro[1], E. Jacquet[1], C. Deranlot[1], P. Bortolotti[1], R. Lebourgois[4], J-C. Mage[1], G. de Loubens[3], O. Klein[3], V. Cros[1] and A. Fert[1]

[1]Unité Mixte de Physique CNRS/Thales and Université Paris-Sud, 1 avenue Augustin Fresnel, Palaiseau, France

[2]Université de Bretagne Occidentale, LMB-CNRS, Brest, France

[3]Service de Physique de l'Etat Condensé, CEA/CNRS, Gif-sur-Yvette, France

[4]Thales Research and Technology, 1 avenue Augustin Fresnel, Palaiseau, France

**Abstract:**

High quality nanometer-thick (20 nm, 7 nm and 4 nm) epitaxial YIG films have been grown on GGG substrates using pulsed laser deposition. The Gilbert damping coefficient for the 20 nm thick films is $2.3 \times 10^{-4}$ which is the lowest value reported for sub-micrometric thick films. We demonstrate Inverse spin Hall effect (ISHE) detection of propagating spin waves using Pt. The amplitude and the lineshape of the ISHE voltage correlate well to the increase of the Gilbert damping when decreasing thickness of YIG. Spin Hall effect based loss-compensation experiments have been conducted but no change in the magnetization dynamics could be detected.

* Contact author :
Abdelmadjid Anane
abdelmadjid.anane@thalesgroup.com

Among all magnetic materials, Yttrium Iron Garnet $Y_3Fe_5O_{12}$ (YIG) has been the one that had the most prominent role in understanding high frequency magnetization dynamics. Because of its unique properties, bulk YIG crystal was the prototypal material for ferromagnetic resonance (FMR) studies in the mid-twentieth century. The attractive properties of YIG include: high Curie temperature, ultra low damping (the lowest among all materials at room temperature), electrical insulation, high chemical stability and easy synthesis in single crystalline form. Micrometer thick films of YIG were first grown using liquid phase epitaxy (LPE)[1], and paved the way for the emergence of a large variety of microwave devices for high-end analogue electronic applications throughout the 1970's [2].

More recently, the interest in emerging large-scale integrated circuit technologies for beyond CMOS applications has fostered new paradigms for data processing. Many of them are based on state variables other than the electron charge and may eventually allow for unforeseen functionalities. Coding the information in a spin wave (SW) is among the most promising routes under investigation and has been referred to as magnonics [3]. Exciting and detecting spin waves has been mainly achieved through inductive coupling with radiofrequency (rf) antennas but this technology remains incompatible with large scale integration[4]. Disruptive solutions merging magnonics and spintronics have been recently proposed where spin transfer torque (STT) and magnetoresistive effects would be used to couple to the SWs. For instance, using a STT-nano-oscillator in a nanocontact geometry, coherent SWs emission in $Ni_{81}Fe_{19}$ (Py) thin metallic layer has been recently demonstrated and probed by micro-focused Brillouin light scattering (BLS) [5].

YIG is often considered as the best medium for SW propagation because of its very small Gilbert damping coefficient ($2\times10^{-5}$ for bulk YIG). Being an electrical insulator, electron mediated angular momentum transfer can only occur at the interface between YIG and a metallic layer. In that context, metals with large Spin Orbit Coupling (SOC) like Pt where a pure spin current can be generated through Spin Hall Effect (SHE) [6] have been used to excite[7] or amplify[8,9] propagating SWs through loss compensation in YIG. Moreover, detection of SW can be achieved using the Inverse Spin Hall Effect (ISHE). In ISHE, the flow of a pure spin current from the YIG into the large SOC metal generates a dc voltage. The ISHE voltage is proportional to the Spin Hall angle ($\theta_{SH}$) and the effective spin mixing conductance ($G_{\uparrow\downarrow}^{eff}$) that is in play in the physics of spin pumping. As for the SOC materials, up to recently, mainly Pt has been used, however it can be observed that other $5d$ heavy metals such as Ta [10], W [11] or CuBi [12] are also very promising.

As the amount of angular momentum transferred from (to) the YIG magnetic film per unit volume scales with $1/t$ (where $t$ is the YIG thickness), it is necessary to reduce the YIG thickness as much as possible while keeping its magnetic properties. Indeed, the threshold current density in the SOC metal for the macrospin mode excitation is expressed as[13] : $j_c = \frac{1}{\theta_{SH}} \cdot \frac{e\alpha f M_s t}{\gamma \hbar}$ (Eq. 1) where $\theta_{SH}$ is the spin Hall angle, $e$ the electron charge, $\gamma$ the gyromagnetic ratio, $f$ the FMR frequency and $M_s$ the YIG's saturation magnetization. Furthermore, a better understanding of the physics involved in spin momentum transfer at the YIG/metal interface would be achieved by reducing the YIG film thickness below the exchange length (~ 10 nm) [14]. Up to now, sub micrometer-thick YIG films have been mainly grown by LPE but the ultimate thickness are around 200 nm[15]. To further reduce the thickness, other growth methods are to be considered. Pulsed laser deposition (PLD) is the most versatile technique for oxide films epitaxy. Several groups have worked on PLD grown YIG[16,17,18,19] but it is only recently that the films quality is approaching that of LPE[20,21].

In this letter, we present PLD growth of ultrathin YIG films with various thickness (20 nm, 7 nm and 4 nm) on Gadolinium Gallium Garnet (GGG) (111) substrates. Structural and magnetic characterizations and FMR measurements demonstrate the high quality of our nanometer-thick YIG films comparable to state of the art LPE films. The growth has been performed using a frequency tripled ($\lambda$ = 355 nm) Nd:YAG laser and a stochiometric polycrystalline YIG pellet. The pulse rate was 2.5 Hz and the substrate-target distance was 44 mm. Prior to the YIG deposition, the GGG substrate is annealed at 700°C under an oxygen pressure of 0.4 mbar. Growth temperature is then set to be 650 °C, and oxygen pressure to 0.25 mbar. After the film deposition, samples are cooled down to room temperature under 300 mbar of $O_2$. The YIG thickness is measured for each sample using X-ray reflectometry which yield a precision better than 0.3 nm. The surface morphology and roughness have been studied by atomic force microscopy (AFM). RMS roughness has been measured over 1 $\mu m^2$ ranges between 0.2 nm and 0.3 nm for all films (Fig 1a). As often with PLD growth, droplets are present on the film surface, here their lateral sizes are below 100 nm and their density is very low (~ 0.1 $\mu m^{-2}$). X-Ray Diffraction (XRD) spectra using Cu K$\alpha_1$ radiation show that the growth is along the (111) direction. Only peaks characteristic of YIG and the GGG substrate are observed (see Fig 1b). The YIG lattice parameter is very close to that of the substrate and can only be resolved eventually at large diffraction angles. For the 20 nm YIG film (Fig 1b,1a), refinement using EVA software on the 888 reflection yields a cubic lattice parameter of 1.2459 nm, to be compared to 1.2376 nm for the bulk YIG[22]. For thinner films (between 4 and 15 nm), it was not possible to distinguish the YIG peaks from those of the substrate (Fig 1d, 1e). This sharp variation of the XRD spectra with respect to the film thickness tends to point towards a critical thickness for strain relaxation. It is however worth noting that the cubic lattice parameter of the 20 nm thick film is larger than the bulk lattice parameter but also of that to the GGG substrate (1.2383 nm). A slight off-stoichiometry (either oxygen vacancies or cation interstitials) is probably at the origin of this observation. Pole figure measurements have been performed to gain insights into the in-plane crystal structure, but it was not possible to resolve, at this stage, the films peaks from the substrate peaks from which we infer that the growth is epitaxial and the film single crystalline.

From SQUID magnetometry with in-plane magnetic field, we measure a magnetization of $4\pi M_s$ = 2100 G ± 50 G at room temperature for both the 20 and 7 nm films. This value is independently confirmed by out-of plane FMR resonance while the tabulated bulk value for YIG is $4\pi M_s = 1760$ G. A similar increase of the PLD grown YIG magnetization have been reported and attributed to an off-stoichiometry[19]. The coercive fields are extremely small, about 0.2 Oe (which is the experimental resolution) and the saturation field is 5 Oe. There is no evidence for in-plane magnetic anisotropy. The overall magnetic signature is that of an ultra-soft material. Note that for the thinnest films (4 nm) we measure a decrease of the saturation magnetization to roughly 1700 G. Finally, we emphasize that the structural and magnetic properties of the samples are well reproducible with respect to the elaboration conditions.

FMR fields and linewidths were measured at frequencies in the range 1-40 GHz using high sensitive wideband resonance spectrometer with a nonresonant microstrip transmission line. The FMR is measured via the derivative of microwave power absorption using a small rf exciting field. Resonance spectra were recorded with the applied static magnetic field oriented in plane. During the magnetic field sweeps, the amplitude of the modulation field was appreciably smaller than the FMR linewidth. The amplitude of the excitation field $h_{rf}$ is about 1 mOe, which corresponds to the linear response regime. A phase-sensitive detector with lock-in detection was used. The field derivative of the absorbed power is proportional to the field derivative of the imaginary part of the rf susceptibility: $\frac{dP_a}{dH} \propto \frac{d\chi''}{dH}$ (where $P_a = \frac{1}{2} \omega \chi'' h_{rf}^2$ and $\chi''$ is the imaginary part of the susceptibility of the uniform mode). Typical resonance curves are plotted in Fig. 2a 2b. In Fig. 2c, we show the frequency dependence of the peak-to-peak linewidth for three different YIG thicknesses, *i.e.*, 20, 7 and 4 nm. As for the 20 nm YIG film, we find a linear dependence of the FMR linewidth with rf frequency while for the thinnest films, we do find an almost linear increase in the low frequency range (< 12 GHz) and then a saturation of the linewidth with frequency. Such qualitative difference depending on the thickness is reminiscent of the qualitative difference discussed earlier in the X-ray diffraction data (Fig 1). The linear dependence of the resonance linewidth is expected within the frame of the Landau-Lifshitz Gilbert equation and allow for a straightforward calculation of the intrinsic Gilbert damping coefficient ($\alpha = 2.3 \cdot 10^{-4}$ for the 20 nm thick film). The zero frequency intercept of the fitting line, usually referred to as the extrinsic linewidth [23,24], is found to be $\Delta H_0$ =1.4 Oe. We emphasize that our value for the intrinsic damping on the 20 nm thick film is among the best ever reported independently of the growth technique and is only outperformed by the 1.3 μm film used by Y. Kajiwara *et al.* [7]. As for the extrinsic damping, our values are still a bit larger than those obtained for 200 nm thick films grown by LPE ($\Delta H_0$= 0.4 Oe) [10]. The saturation of the linewidth with increasing excitation frequency observed for thinnest films (*t* = 7 and 4 nm) is usually ascribed to two-magnons scattering due to the interfaces [25]. An estimation of the intrinsic damping in such thin films is thus not correct. Nevertheless, and only for the sake of comparison, considering frequencies under 6 GHz, we can roughly estimate the low frequency Gilbert damping to be $1.6 \cdot 10^{-3}$ for the 7 nm and $3.8 \cdot 10^{-3}$ the 4nm YIG films. However, it is worth mentioning that for those two thinnest films; samples sliced from the same substrate can give different linewidths (up to a factor of 3) with for some of them up to 2 absorptions lines. This observation points to a slight lateral non-homogeneity in the chemical composition[24]. The data presented in figure 2 are those of the best samples showing a single absorption line. For the 20 nm thick films, all samples have only one resonance line and the dispersion of linewidths is within 5%.

In order to characterize the conversion of propagating SWs in YIG into a charge current in a normal adjacent metal with large SOC, we perform ISHE detection of SW with the strip geometry used by Chumak et al.[26]. In our sample design (see Fig 3a), SWs excitation is achieved using a patterned 100 µm wide Au stripline antenna whereas the ISHE voltage is measured on a 13 nm thick Pt strip (0.2 mm x 5 mm) located at 100 µm away from the Au stripe and parallel to it. The metallic Pt strip is deposited using dc magnetron sputtering and lift-off. Prior to the Pt deposition, an in-situ $O_2$/Ar-plasma is used to remove the photo-resist residues and increase the ISHE voltage. This cleaning step has been shown recently to improve the ISHE signal by one order of magnitude[27,28]. Measurements of the ISHE signal is performed either using a lock-in (with a 5 kHz TTL modulation of the rf power) or a nano-voltmeter. In order to increase the output ISHE signal, we chose a specific configuration with a magnetic field at 45° from to the SW propagation direction (cf Fig. 3a). This configuration has the advantage of providing a good coupling of the YIG film to the rf field under the antenna while still having a significant spin polarization ($\sigma$) that is orthogonal to the measured ISHE electrical field. We should point out that our choice of magnetic field direction implies that the propagating SWs are neither Damon-Eshbach modes where **k** ⊥ **M** nor backward volume modes where **k** // **M**.

In Fig 3b, we display the ISHE voltage (without any geometrical correction) as a function of the in plane magnetic field measured on a 20 nm thick YIG under a 10 mW rf excitation at 1 GHz. The sign of voltage peak (occurring at magnetic fields that resonantly excite the magnetization) reverses when the applied field is reversed as expected from the ISHE symmetry [29] : $V^{ISHE} \propto J_s \times \sigma \cdot u_y$ where $J_s$ is the pure spin current that flows through the YIG/Pt interface, $\sigma$ is the spin polarization vector parallel to the dc magnetization direction and $u_y$ is the unit vector parallel to the Pt strip. A close-up on the ISHE signal lineshape for the different thicknesses is plotted in Fig. 3c 3d 3e. In accordance with FMR study, the linewidth increases with decreasing YIG thickness. The spectral lineshapes are almost symmetrical confirming previous reports on YIG films grown using LPE [30]. The ISHE maximum voltage decreases when the film thickness is decreased. Going from 20 nm to 4 nm this decrease is as large as two orders of magnitudes and correlates to the increase of the Gilbert damping. A decrease of the ISHE signal when increasing damping is expected. Indeed, as we are in the weak excitation regime, the ISHE voltage is expected to scale with $1/\alpha^2$, see for instance supplementary materials of Ref [7]. If we consider the Gilbert damping obtained from FMR on the bare YIG samples, a more dramatic decrease of the ISHE signal than the one observed is expected. In fact here, one should consider the effective damping of the Pt/YIG stack as spin-pumping will increase the YIG effective damping under the Pt strip. We should again point out that our measurement geometry relies on propagating SWs and therefore they are subject to an exponential decay with distance. The length scale of this exponential decay is different for the three thicknesses owing to the difference in the damping parameter; hence a quantitative interpretation of the voltage amplitude is to be avoided here.

We thus succeeded to grow high quality ultrathin YIG film and demonstrated that an efficient spin angular momentum transfer from the YIG film toward the Pt layer. The next objective is to induce an modification on the effective YIG damping coefficient via interfacial spin injection using SHE, as it has been reported in Py/Pt[31,32] system. Using YIG/Pt, Kajiwara *et al.* have shown that even without rf excitation, spin injection induced by a dc current in Pt can generate propagating SWs; the threshold current density has been estimated to be 4.4 x $10^8$ A.m$^{-2}$. Such unexpectedly low value has been attributed to the presence of an easy–axis surface anisotropy[13]. Hence, we have performed experiments where a large dc bias current is applied on the Pt electrode while pumping spin waves in the 20 nm thick YIG film with TTL modulated rf excitation field. The $V^{ISHE}$ linewidth is measured at the TTL modulation frequency using a lock-in. We expected to increase or decrease this linewidth depending on the dc current polarity but we have not been able to see any sizable effect even for current densities as large as 6 x $10^9$ Am$^{-2}$, within a 0.2 Oe resolution, the measured linewidth remains absolutely unchanged. Theory predicts that the threshold dc current for the onset of magnetic excitations scales with the Gilbert damping and the thickness of the ferromagnetic insulator (Eq. 1), the smaller the $\alpha \times t$ product the lower the threshold current for SW excitation is. In Kajiwara *et al.*'s experiment $\alpha \times t = 0.087$ nm; in our case $\alpha \times t \sim 0.01$ nm (considering the spin pumping contribution to the damping). We therefore applied a dc current that is roughly 2 orders of magnitude larger than the expected threshold current. One possible explanation is that in our films, surface anisotropy is absent and therefore the pumped angular momentum is spread over many excitations modes[33]. Further investigations are under progress to clarify this point.

In summary, we have fabricated PLD grown YIG on GGG (111) substrates with thickness as low as 4 nm. We present here a comparative study on three different thicknesses: 4, 7 and 20 nm. The Gilbert damping coefficient for the 20 nm thick films is 2.3 x $10^{-4}$ which is the lowest value reported for sub-micrometric thick films. We demonstrate ISHE detection of SWs for ultra thin YIG film. The amplitudes of the ISHE voltage correlates well to the increase of the Gilbert damping when decreasing thickness. Owing to extremely low $\alpha \times t$ product of the 20 nm film that is almost 10 times smaller than the one reported by Kajiwara *et al.* we expected to observe compensation of the damping by spin current injection through the SHE but our preliminary results on the $V^{ISHE}$ linewidth did not reveal any effect on the magnetization dynamics.

**Acknowledgement:**

This work has been supported by ANR-12-ASTR-0023 Trinidad

Figure Captions :

Figure 1:

(color online). (a) 1 μm x 1μm AFM surface topography of a 20nm YIG film on GGG (111) (RMS roughness=0.23 nm). (b) XRD : θ-2θ scan of the same film using Cu-Kα$_1$ radiation. This spectrum shows that the growth is along the (111) direction with no evidence of parasitic phases. The YIG peak are best resolved at high diffraction angle, a zoom around the 888 GGG peak is shown in panel (c).

(c) (d) (e) XRD diffraction spectrum centered on GGG (888) reflection for different YIG thicknesses: 20, 7 and 4 nm. For the 2 thinnest films, the YIG peak is masked by the substrate peak. The cubic lattice parameter of the 20 nm thick YIG film (obtained from 888 peak) is slightly larger than that of the substrate (a$_{YIG}$=1.2459 nm, a$_{GGG}$=1.2383 nm)

Figure 2:

(color online) (a) (b). FMR absorption derivative spectra of 20 and 4 nm thick YIG films at an excitation frequency of 6 GHz.

(c) rf excitation frequency dependence of FMR absorption linewidth measured on different YIG film thicknesses with an in-plane oriented static field. The black continuous line is a linear fit on the 20 nm thick film from which a Gilbert damping coefficient of 2.3 x 10$^{-4}$ can be inferred ($\Delta H^{p-p} = \Delta H_0 + \alpha \frac{4\pi}{\gamma \sqrt{3}} f$). The damping of the 7 nm and 4 nm films is significantly larger but must off all the frequency dependence is not linear (see text for discussion).

Figure 3:

(a) Schematic illustration of the experimental setup: spin waves are excited through the YIG waveguide with a microstrip antenna. The detection is performed by measuring ISHE voltage dc signal on a ~200μm wide Pt stripe. (b) External dc magnetic field dependence of ISHE voltage measured on Pt electrode showing the polarity inversion when the magnetic field is reversed. The peaks occur at the FMR conditions as verified through S$_{11}$ spectroscopy on the Au antenna (not shown here).

(c). (d). (e). ISHE voltage for different YIG film thicknesses around the resonance magnetic field; microwave frequency is *f*=3GHz. The peaks can be fitted to a lorentzian shape and the extracted linewidth are respectively: 19.2 , 13.2 and 4.6 Oe. The dramatic increase of the ISHE signal with thickness is discussed in the text.

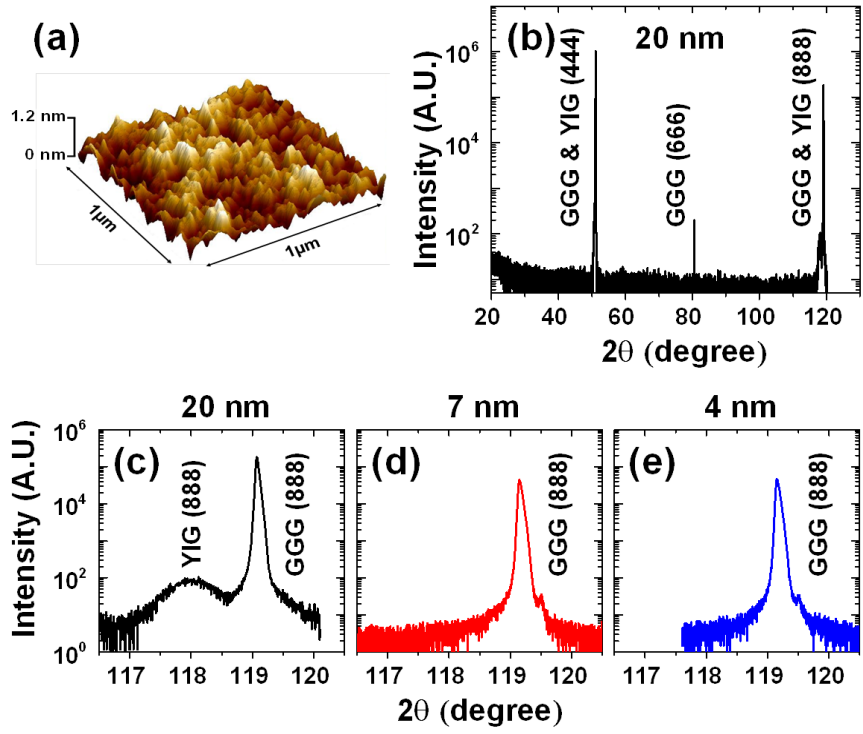

Fig 1

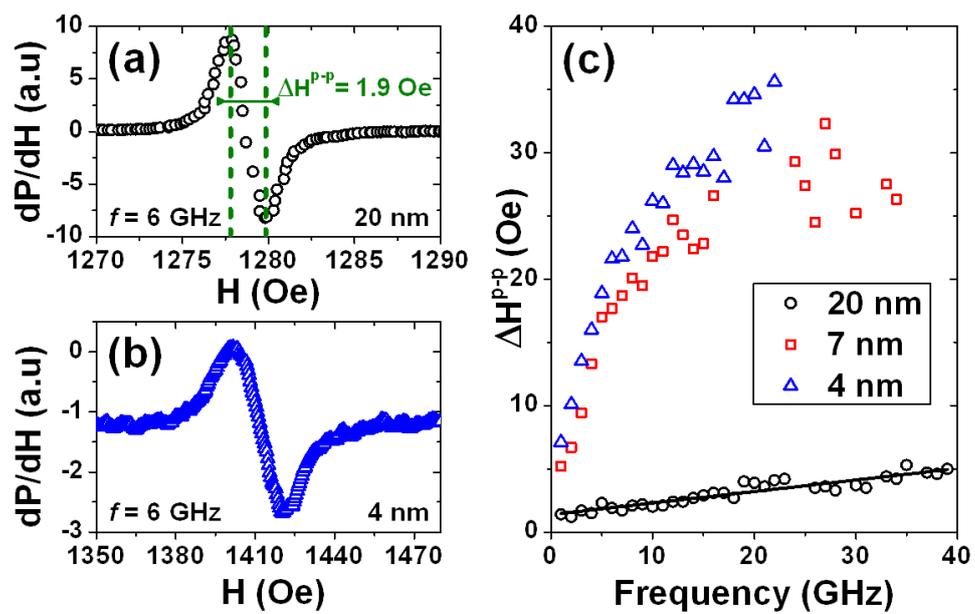

Fig 2

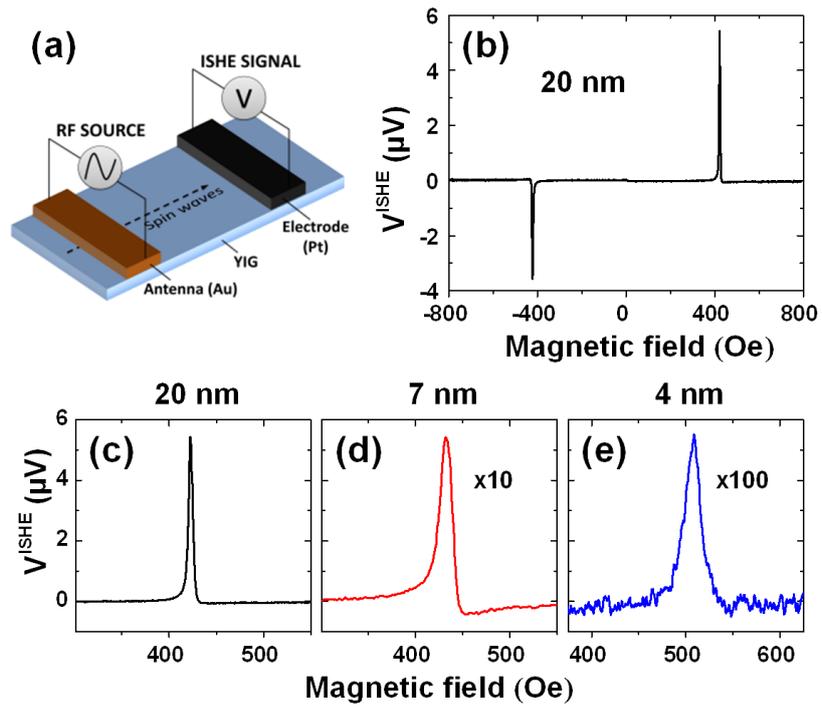

Fig 3